\documentclass[11pt,aps,prd,notitlepage]{revtex4-1}

\usepackage[english]{babel}
\usepackage[font={small},flushleft,indent]{caption}

\usepackage{amsmath,float} 
\usepackage{graphicx}
\usepackage[colorlinks=true, allcolors=blue]{hyperref}
\usepackage{subcaption}
\usepackage{placeins}
\usepackage{overpic}

\begin{document}

\title{Investigating non-Keplerian motion in flare events with astrometric data}
\author{Fengting Xie}
\email{xiefengting@stu.cqu.edu.cn} 
\author{Qing-Hua Zhu}
\email[Corresponding author: ]{zhuqh@cqu.edu.cn} 
\author{Xin Li}
\email{lixin1981@cqu.edu.cn} 
\affiliation{School of Physics, Chongqing University, Chongqing 401331, China}

\begin{abstract}
The GRAVITY interferometer has achieved microarcsecond precision in near-infrared interferometry, enabling the tracking of flare centroid motion in the strong gravitational field near the Sgr A*. It might be promising to serve as a unique laboratory for exploring the accretion matter near black holes or testing Einstein's gravity. Recent studies debated whether there is a non-Keplerian motion of the flares in the GRAVITY dataset. This motivates us to present a comprehensive analysis based on error estimation under the Bayesian framework. This study uses astrometric flare data to investigate the possibility that the flares exhibit deviations from the circular Keplerian motion.
We analyze both averaged and individual flare data, modeling the hotspot with either circular orbits parameterized by a non-Keplerian correction or planar geodesic orbits. It is confirmed that the astrometric data favor the circular orbits over non-circular ones, with the orbital circularity parameter of $\gamma = 0.99_{-0.10}^{+0.07}$. Our results show that the joint posteriors for black hole mass and non-Keplerian parameter are negatively correlated. Fixing the mass to be its established value yields a non-Keplerian parameter of $\omega/\omega_k = 1.45^{+0.35}_{-0.38}$, at approximately the 1$\sigma$ level. The statistical significance is insufficiently high, and the conclusion is found to be sensitive to the presence of correlations in the astrometric data, which might originate from  the non-uniform $u$-$v$ coverage in interferometer measurements. In this sense, the current data might be insufficient to draw a definitive conclusion regarding the presence of non-Keplerian motion. Future improvements in astrometry precision might enable stronger constraints on the kinematical behavior of the flares. 

\end{abstract}

\maketitle
\section{Introduction}
In weak field regime of gravity, pioneers' studies have confirmed the validity of general relativity to a certain extent \cite{2003:GRCassini,Confrontation:GRExperiment,abuter2018redshift,abuter2020detection,2019:redshifts0-2,2017:GReffectS2}, including classical experiments such as the Shapiro time delay \cite{2003:GRCassini}, constraints on parametrized post-Newtonian (PPN) parameters \cite{Confrontation:GRExperiment}, as the gravitational redshift and Schwarzschild precession observed in the orbit of the S2 star \cite{abuter2018redshift,abuter2020detection}. The relativistic effect is necessary because the observations can not be fully explained by Newtonian gravity alone.

In the strong gravitational field regime,  the closest known supermassive black hole (SMBH) at the center of our galaxy, namely, Sagittarius A* (Sgr A*),  might serve as a unique laboratory for exploring the accretion matter near the black holes and testing Einstein's gravity \cite{Narayan:2023ivq,Levis:2023tpb,Galishnikova:2022mjg,genzel2021forty}. The supermassive black hole with a mass of $M=4.3\times10^6M_\odot$  \cite{GRAVITY:2021xju} and a distance of $D=8.178\text{kpc}$ \cite{RevModPhys.82.3121} has been the focus of continuous observation over the past decades \cite{genzel2021forty}. In 2022, the Event Horizon Telescope (EHT) Collaboration released the first image of Sgr A* \cite{EventHorizonTelescope:2022wok,milkywayimage}, reconstructed using very-long-baseline interferometry (VLBI). The image revealed the potential shadow of the black hole and provided a support for the validity of general relativity at the event horizon scale \cite{Perlick:2021aok,Vagnozzi:2022moj,Zhang:2024jrw,Chen:2023wzv,Kuang:2024ugn,Li:2020drn,Chang:2021ngy,Zhu:2022shb,He:2024qka}.

With the advancement of observational techniques, the GRAVITY Collaboration has achieved microarcsecond precision in near-infrared interferometry, enabling tracking of flare centroid motion in the strong gravitational field near the Sgr A* \cite{GRAVITYFirstlight,abuter2018detection,abuter2023polarimetry}. The GRAVITY interferometer, installed at the Very Large Telescope (VLT), is dedicated to high-precision monitoring of the Galactic Center black hole and reported its detection of flares from Sgr A* in 2018 \cite{abuter2018detection}. These data include both astrometry and polarization measurements. Subsequently, the GRAVITY Collaboration updated the dataset of the Sgr A* flares, reporting four flares with astrometric measurements and six with polarimetric data \cite{abuter2023polarimetry}.  Notably, two of flares were well covered in both domains. All astrometric measurements revealed a consistent clockwise motion of the flare centroids on the sky with periods of approximately one hour, accompanied by full rotations of the polarization vector at the same time. The astrometry of these flares are located very close to the Galactic Center, within a few Schwarzschild radii, implying that general relativity's strong-field predictions could be tested. 

The flares might  physically originate from the dynamics of accretion matter in the vicinity of the center black hole \cite{2023:reconnectionflares,2023Magneticflares}. Its phenomenological behavior could be described as a compact blob of plasma orbiting near the innermost stable circular orbit (ISCO), referred as to hotspot model \cite{bao1992signature,Li:2014coa,IntroductionHotspotModel,2023:GReff-flare,polarizationhotspotintoKerrBH,flaregeodesichotspotKerr,hotspotscalarizedReissner,Rosa:2024bqv,Rosa:2022toh,Tamm:2023wvn,Rosa:2023qcv,zhu2025observational,Wei:2024cti}. 
Based on GRAVITY datasets,  Refs. \cite{orbitalorpatternmotion,antonopoulou2024parameter} explored various kinematic scenarios for flare motion. Notably, Ref.~\cite{antonopoulou2024parameter} showed that super-Keplerian orbits often provide a better match to the observations. 
Based on the astrometric data, GRAVITY Collaboration  reported that while centroid motion is generally consistent with a circular orbit, the observed orbital period is shorter than the predicted Keplerian period \cite{baubock2020modeling}. Further studies by Ref.~\cite{yfantis2024hot} provided additional support for the super-Keplerian interpretation. However, Refs. \cite{abuter2018detection,abuter2023polarimetry} still emphasized that the flare motions remain consistent with a standard Keplerian model. 

Motivated by the previous studies, this paper analyzes the flare data released by the GRAVITY Collaboration in 2023, and performs hotspots' orbit fitting for the four flares in the Bayesian parameter estimation framework. The astrometric data of the four flare events can also be averaged, due to the similarity between the flares \cite{abuter2023polarimetry}. We will separately analyze the averaged and individual data with the error estimation and provide a comparison of the results. We also study non-circular motion of the flares, exploring its statistical connection to the indications of the non-Keplerian motion. Based on the astrometric data, we will examine whether the flares show statistical significance of the circular non-Keplerian motion around Sgr A*.


The rest of the paper is organized as follows. In Section \ref{methods}, we present the observed data, describe our ray-tracing method for modeling the emission from the hotspot, and outline the setup of parameter estimation in Bayesian framework. Section \ref{results} details the fitting results using both the averaged and individual flare data, under different orbit configurations. In Section \ref{conclusion}, we summarize our findings and provide a discussion of the implications.

\section{Datasets and Methods}\label{methods}

In this section, we describe the methodology for modeling and fitting the apparent tracks of flares, including ray tracing and parameter estimation via Markov Chain Monte Carlo (MCMC) sampling. 
We adopt the hotspot model, where observed flares originate from a bright spot orbiting on the surface of accretion disks \cite{genzel2010galactic,bao1992signature,zhu2025observational}. 
Because smaller hotspot size tend to yield better fitting results \cite{baubock2020modeling}, we consider point-like hotspots in this study. Due to the limited amount and low precision of the available data, it seems not possible to constrain the spin of the black hole \cite{abuter2018detection,abuter2023polarimetry,yfantis2024hot}.  We thus consider the Schwarzwald black hole in the subsequent sections.

\subsection{GRAVITY dataset}

The data used in this study were reported by the GRAVITY Collaboration in 2023 \cite{abuter2023polarimetry} including 6 with polarimetric data and 4 with astrometric data. All astrometric measurements revealed a consistent clockwise motion of the flare centroids on the sky with periods of approximately one hour, accompanied by full rotations of the polarization vector at the same time. The astrometry of these flares are located very close to the Galactic Center, within a few Schwarzschild radii. Recent studies suggest that the hotspot may exhibit super-Keplerian motion around Sgr A* \cite{baubock2020modeling,antonopoulou2024parameter,yfantis2024hot}, which is a discrepancy with the conclusion in Ref.~\cite{abuter2018detection,abuter2023polarimetry}. It is noted that GRAVITY Collaboration fitted the flare motion with a Schwarzschild orbital model using averaged data \cite{abuter2023polarimetry}. While averaging data benefits from canceling systematic biases in the measurements, there is still concern that the use of averaged data may smooth over important details of the flare. In this paper, we therefore study the flare motion using both the averaged astrometric data and the astrometric data from individual four flare events. 

\subsection{Hotspot orbital configuration and ray tracing method}

The hotspot model and ray-tracing simulation are conducted in the spherical black hole spacetime with the metric as follows,
\begin{equation}
    \textrm{d}s^2=-f(r)\textrm{d}t^2+\frac{\textrm{d}r^2}{f(r)}+r^2(\textrm{d}\theta^2+\sin\theta^2\textrm{d}\phi^2)~.
\end{equation}
The Schwarzschild black hole is considered in this study, where $f(r)=1-2M/r$, because current GRAVITY dataset seems difficult to constrain the black hole spin \cite{abuter2018detection,yfantis2024hot}.

The hotspots in circular orbits could be the most promising models, since it corotates with the accretion disks as suggested in Refs.~\cite{abuter2018detection,abuter2023polarimetry}.
Angular velocity of hotspots induced by pure gravitational force is $\omega_K\equiv u^{(\text{cir}),\phi}/ u^{(\text{cir}),t}=(m/r_0^3)^{1/2}$. It is consistent with the results in Newtonian gravity, referred as to Keplerian  motion.
In the case of deviations from Keplerian motion, the four-velocity of hotspots can be rewritten in the form of
\begin{equation}
    u^{(\text{dK})}=\frac{1}{\sqrt{f(r_0)- r_0^2\omega^2}}\partial _t\pm\frac{\omega}{\sqrt{f(r_0)- r_0^2\omega^2}}\partial_\phi
\end{equation}
where the parameter $\omega/\omega_K>1$ denotes the super-Keplerian motions, and $\omega/\omega_K<1$ denotes the sub-Keplerian motions.
Beside above circular motion, there is also a possibility that the indication of non-Keplerian motions might originate from non-circular orbits. Therefore, we consider the hotspot moving along planar geodesic orbits. In this case, the initial four-velocity can be written as
\begin{equation}
    u^{(v)}_0=\sqrt{\frac{1}{f(r_0)}\left(1+\frac{L}{r_0^2}\right)}\partial_t+\frac{L}{r_0^2}\partial_\phi~,
\end{equation}
where $L$ is the angular momentum, and $r_0$ is the initial radius of the hotspot orbits, 
In the following analysis, we consider two orbital models: the hotspot is set to move either along a circular orbit or along a geodesic, both confined to the equatorial plane.

Because the flares are much closer to the center black hole, the bending of light caused by gravity can not be neglected in observations \cite{orbitalorpatternmotion,abuter2018detection}.
The conventional ray-tracing algorithm, often referred to as “backward ray tracing", traces light rays backward from each pixel on the image plane to their potential emission sources \cite{vincent2011gyoto, dexter2009fast,dexter2016public}. This approach is effective for the sources like accretion disks, but it becomes computationally expensive for small or point-like sources, as most light rays might not reach the emission region. We adopt the ray-tracing code developed in Ref.~\cite{zhu2025observational}, which determines the light paths based on the given locations of the emission source and the observer. This approach can significantly reduce the computational cost for the point-like sources. Here, to fit astrometric data of the flares, the apparent tracks of hotspots on observers' sky are given by the flux-weighted centroid position combining the hotspots' primary and secondary images \cite{baubock2020modeling,zhu2025observational}. 

\subsection{Fitting procedure}

\begin{figure}
    \centering
    \includegraphics[width=0.9\linewidth]{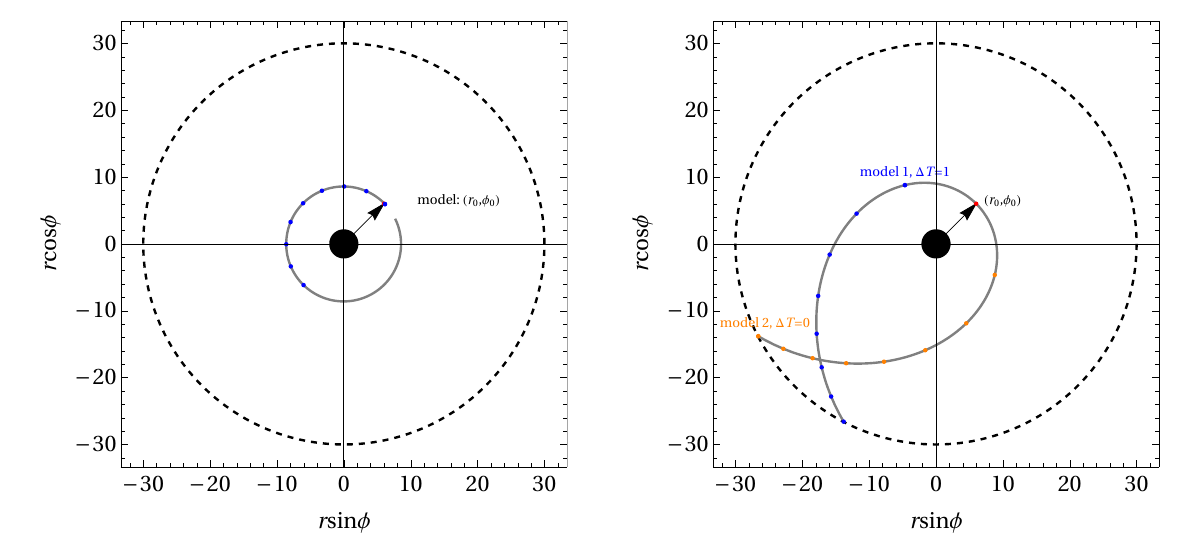}
    \caption{Schematic diagram for illustrating the parameters of hotspots in circular orbits (left panel) and planar geodesic orbits (right panel). The position $(r_0,\phi_0)$ is the initial position for the circular orbits, and is the innermost point for the planar geodesic orbits. The $\Delta T$ formulates the relative distance to the innermost point. The dashed circles represent the boundary of the positions in $(r,\phi)$ that we studied. }
    \label{F1}
\end{figure}

We perform the Markov Chain Monte Carlo (MCMC) sampling using the EMCEE method to explore the parameter space in the Bayesian framework.  
The parameters for circular orbits and planar geodesic orbits are presented as follows,
\begin{subequations}
    \begin{eqnarray}
    \Theta_\text{cir}&=&(M/M_\odot,\theta_\text{inc},r_0/M,\phi_0,\text{PA},\omega/\omega_k)~, \label{cir}\\
    \Theta_\text{geo}&=&(M/M_\odot,\theta_\text{inc},r_0/M,\phi_0,\text{PA},L/M^2,\Delta T)~, \label{geo}
\end{eqnarray} \label{model}
\end{subequations}
where $M$ is the black hole mass, $\theta_\text{inc}$ is the inclination angle of the orbit plane, PA is position angle of the orbit plane projected on the observer's sky, and $\omega/\omega_K$ is non-Keplerian parameter. The reference positions $r_0$, $\phi_0$ and $\Delta T$ for our models are illustrated in the schematic diagram in Fig.~\ref{F1}. We consider the planar geodesic orbits that can reach the innermost point at least once, because the astrometric data have shown a consistent clockwise motion of the flare centroids on the sky \cite{abuter2023polarimetry}.

The astrometric data of the flares are given in the form of
\begin{equation}
    \Xi=(X_i,Y_i)~,\label{data}
\end{equation}
where $i[=1,2,...,N]$ is sequential index of data points and $N$ is number of data points.
With the model parameters in Eqs.~(\ref{model}) and dataset in the form of Eq.~(\ref{data}), the log-likelihood function is can be given by
\begin{equation}
    \mathcal{L}(\Theta_\text{model})=-\frac{1}{2}\sum_{i=1}^N\left(\left(\frac{X_i-X_{\text{model}}(t_i)}{\sigma_{X_i}}\right)^2+\left(\frac{Y_i-Y_{\text{model}}(t_i)}{\sigma _{Y_i}}\right)^2\right)~, \label{like1}
\end{equation}
where $X_{\text{model}}(t_i)$ and $Y_{\text{model}}(t_i)$ represent model predictions at time $t_i$ presented in Eqs.~(\ref{model}).  The uncertainties $\sigma_{X_i}$ and $\sigma_{Y_i}$ follow the values reported in Ref.~\cite{abuter2023polarimetry}. 
We run MCMC sampling using 100 chains, each with 2000 iterations for circular orbits, and 3000 iterations for planar geodesic orbits. And flat priors are adopted for all the parameters.
\smallskip 

\section{Analysis of the fitted results}\label{results}

To investigate the kinematic behavior of the flares, we model the apparent hotspot tracks using ray-tracing simulation with given orbit configuration, and subsequently conduct Bayesian parameter estimation. The best-fit parameters were given by globally optimal ones in the samples. The resulting constraints and corresponding uncertainties on orbital radius, inclination, and other relevant parameters are presented. Physically, we focus on whether the orbits show significant deviations from the circular Keplerian motions. Unless otherwise specified, we fix the distance to be black hole $D=8.277$kpc \cite{GRAVITY:2021xju} and adopt the likelihood function in Eq.~(\ref{like1}) with the parameters for circular orbits and planar geodesic orbits defined in Eqs.~(\ref{model}).

\subsection{Hotspots in circular orbits}

\begin{figure}
    \centering
    \begin{overpic}[width=0.8\textwidth]{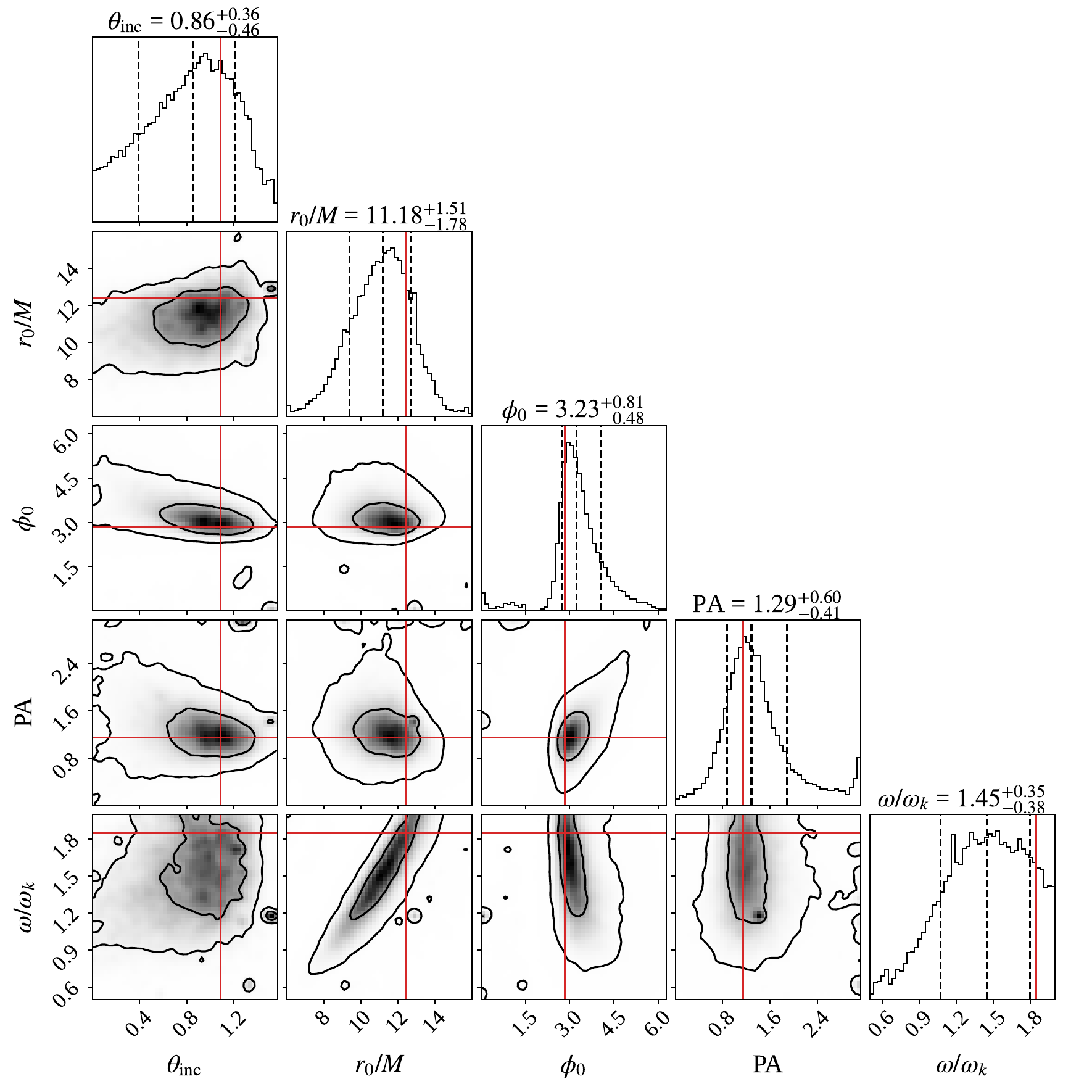} 
     \put(61,54){\includegraphics[width=0.37\textwidth]{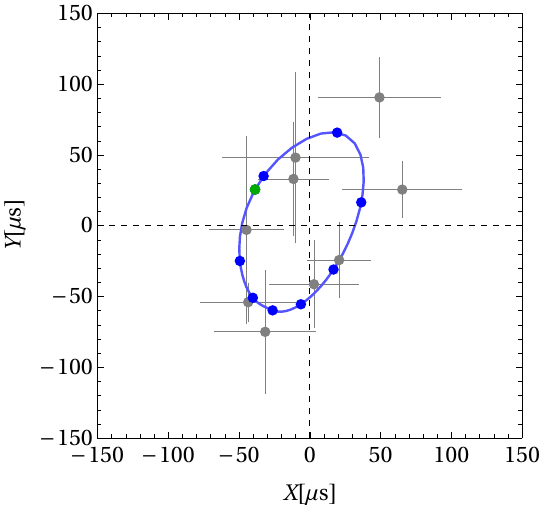}}
    \end{overpic}
    \caption{Bottom-left panel: the posteriors of model parameters obtained from MCMC sampling for the circular orbit model.  The red cross marks the globally best-fit parameters. 
Top-right panel: observed flare
centroids (gray points) and the best-fit track (blue points) of a circularly orbiting hotspot in the equatorial plane.  Here, the black hole mass  $M$ is fixed at $4.3 \times 10^6 M_\odot$ \cite{GRAVITY:2021xju}.  
 We have $\chi^2_\text{eff}=0.29$ for the best-fit parameters.}
    \label{23avg_fixed_mass-diagram_}
\end{figure}

\begin{figure}
    \centering
    \begin{overpic}[width=0.8\textwidth]{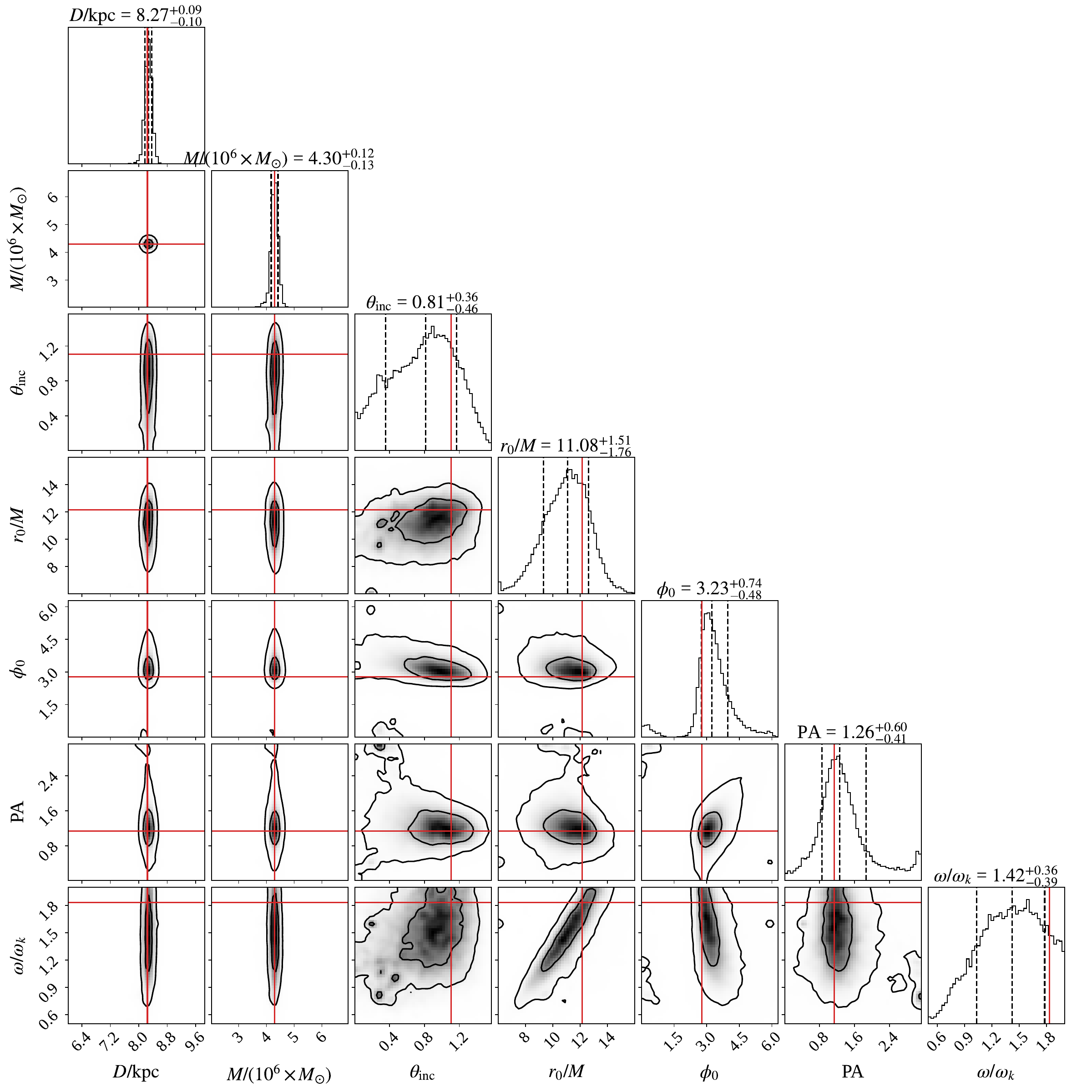} 
     \put(61,54){\includegraphics[width=0.37\textwidth]{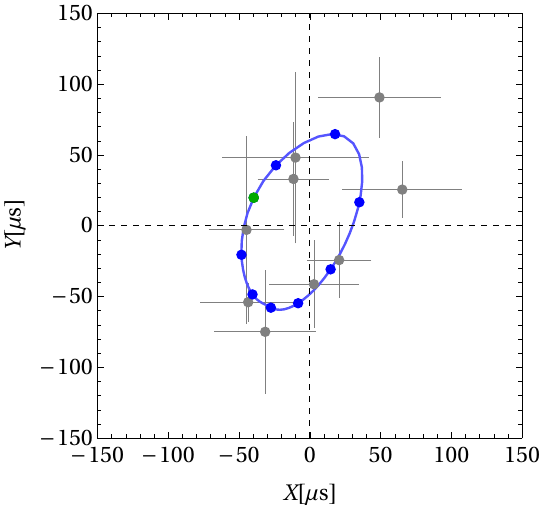}}
    \end{overpic}
    \caption{Bottom-left panel: the posteriors of model parameters obtained from MCMC sampling for the circular orbit model.  The red cross marks the globally best-fit parameters. 
Top-right panel: observed flare
centroids (gray points) and the best-fit track (blue points) of a circularly orbiting hotspot.  We use the Gaussian priors on the black hole mass $M=(4.297 \pm 0.012) \times 10^6M_\odot$ and distance $D=(8.277 \pm 0.009) \text{kpc}$ reported in Ref.~\cite{GRAVITY:2021xju}. The prior widths are set to $10\sigma$.  We have $\chi^2_\text{eff}=0.35$ for the best-fit parameters.}
    \label{23avg_fixed prior-diagram_}
\end{figure}

We begin our analysis by considering the case in which the hotspot moves along a circular orbit confined to the equatorial plane. 
\begin{figure}
    \centering
    \begin{overpic}[width=.9\textwidth]{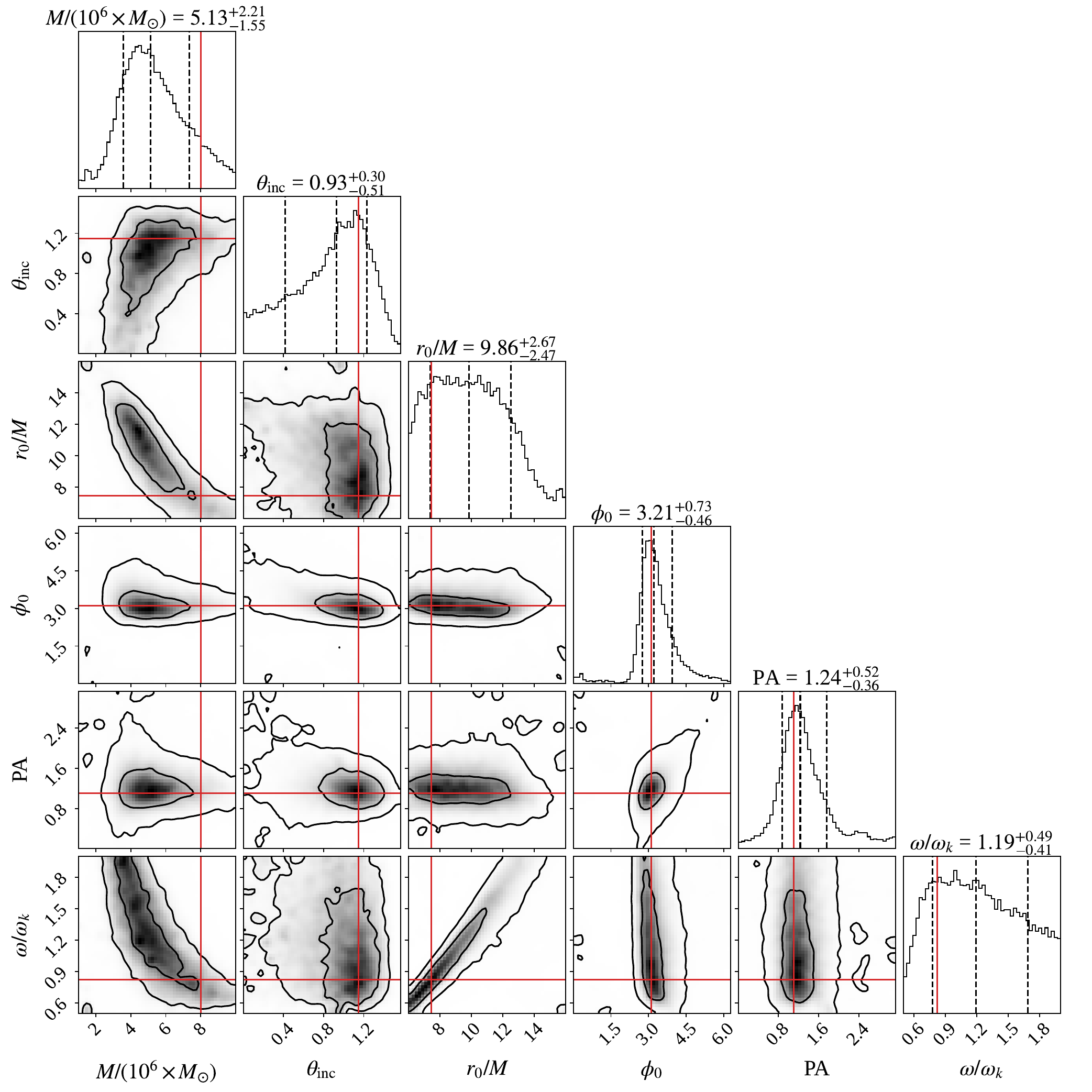} 
     \put(55,57){\includegraphics[width=0.4\textwidth]{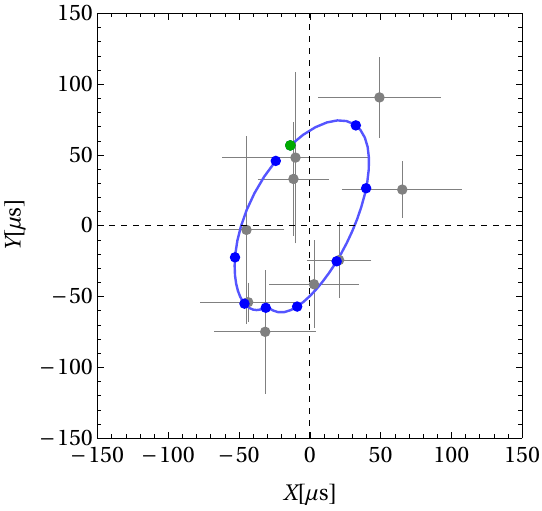}}
    \end{overpic}
    \caption{Bottom-left panel: the posteriors of model parameters obtained from MCMC sampling for the circular orbit model in the equatorial plane. The red cross marks the globally best-fit parameters. Top-right panel: observed flare centroids (gray points) and the best-fit track (blue points) of a circularly orbiting hotspot in the equatorial plane. We have $\chi^2_\text{eff}=0.18$ for the best-fit parameters.
    }
    \label{23avg_diagram_}
\end{figure}
As shown in bottom-left panel of Fig.~\ref{23avg_diagram_}, we used the averaged flare data reported by GRAVITY (2023) \cite{abuter2023polarimetry} and performed parameter estimation for our models. In this case, the black hole mass was treated as a free parameter. The resulting angular velocity is $\omega=1.13^{+0.54}_{-0.42}$, which does not provide conclusive evidence for super-Keplerian motion of the hotspot.  This conclusion is consistent with that of the Ref.~\cite{abuter2023polarimetry}. Furthermore, the inferred mass and orbital radius exhibit a negative correlation in  2D contours, which also agrees with the findings of Ref.~\cite{abuter2023polarimetry}.
In top-right panel of Fig.~\ref{23avg_diagram_}, we shows the best-fit orbit of the hotspot. Our results differ from those reported by the GRAVITY Collaboration (2023) \cite{abuter2023polarimetry}, where the best-fit track looks like a circle. We will discuss this discrepancy in Sec.~\ref{conclusion}. To quantify the goodness of fit, we consider the effective $\chi^2$ as follows,
\begin{equation}
    \chi^2_\text{eff}=-\frac{2}{2N-N_\text{d.o.f}}\mathcal{L}(\Theta_\text{model})~. \label{chieff}
\end{equation}
For the results shown in top-right panel of Fig.~\ref{23avg_diagram_}, we have $\chi^2_\text{eff}=0.18$. It yields a lower effective reduced value $\chi^2_{\rm eff}$, compared to previous studies \cite{abuter2023polarimetry,yfantis2024hot}. From the posteriors in Fig.~\ref{23avg_diagram_}, the  median value of the inferred black hole mass seems larger than the well-established value \cite{2017:GReffectS2,abuter2018redshift,GRAVITY:2021xju}. We therefore fixed the black hole mass $M=4.3\times10^6 M_\odot$ \cite{GRAVITY:2021xju} to preform parameter estimation, which is shown in the bottom-left panel of Fig.~\ref{23avg_fixed_mass-diagram_}. In this setup, the inferred non-Keplerian parameter is shown to be $\omega/\omega_k=1.45^{+0.35}_{-0.38}$, which merely exceeds 1$\sigma$ within the uncertainty range. It indicates that the  presence of super-Keplerian motion still remain insignificant. From the top-right panel of Fig.~\ref{23avg_fixed_mass-diagram_}, the tracks based on the best-fit parameters still agrees well with the observed data, where we have the reduced $\chi^2_\text{eff}(=0.29)$. 
The proper Bayesian approach might employ the priors established from a more accuracy observation, such as the stellar orbits. Therefore, we again use the Gaussian priors based on the black hole mass $M=(4.297 \pm 0.012) \times 10^6M_\odot$ and distance $D=(8.277 \pm 0.009) \text{kpc}$ reported in Ref.~\cite{GRAVITY:2021xju} to perform the parameter estimation. The contour plots for the posteriors are presented in Fig~\ref{23avg_fixed prior-diagram_}. The posteriors for the mass and distance remain unchanged from their priors, indicating the flare astrometric data provides no additional constraint. The posteriors of the remaining parameters are nearly identical in Fig.~\ref{23avg_fixed prior-diagram_} and Fig.~\ref{23avg_fixed_mass-diagram_}. It suggests that the results in Fig.~\ref{23avg_fixed_mass-diagram_} are robust.
We summarize the posterior median values and uncertainties for all parameters in Tab.~\ref{T1}. The results in Fig.~\ref{23avg_fixed prior-diagram_} are omitted from Tab.~\ref{T1} as they are consistent with the fixed black hole mass case.
\begin{table}
    \centering
        \caption{Posterior median values and uncertainties of the parameters modeled by hotspot circular orbits with the averaged astrometric data.}
    \begin{tabular}{c|ccccccc}
    \hline\hline
         Parameter  & $M/(10^6\times M_\odot)$ & $\theta_\text{inc}$   & $r_0/M$   & $\phi_0$   & PA & $\omega/\omega_k$ \\
         \hline
         Free mass  & $5.13^{+2.21}_{-1.55}$    & $0.93^{+0.30}_{-0.51}$   & $9.86^{+2.67}_{-2.47}$                              & $3.21^{+0.73}_{-0.46}$    & $1.24^{+0.52}_{-0.35}$   & $1.19^{+0.49}_{-0.41}$  \\
         Fixed mass & $4.30$ \cite{GRAVITY:2021xju}                      & $0.96^{+0.36}_{-0.46}$   & $11.18^{+1.51}_{-1.78}$                             & $3.23^{+0.81}_{-0.48}$    & $1.29^{+0.60}_{-0.41}$   & $1.45^{+0.35}_{-0.38}$  \\
         \hline
    \end{tabular}
    \label{T1}
\end{table}

\smallskip

\subsection{Analyzing individual astrometric flare data with the circularly orbital hotspots}

\begin{figure}
    \centering
    
      \begin{subfigure}[b]{0.49\textwidth}
      \caption{May 27, 2018}
        \begin{overpic}[width=\textwidth]{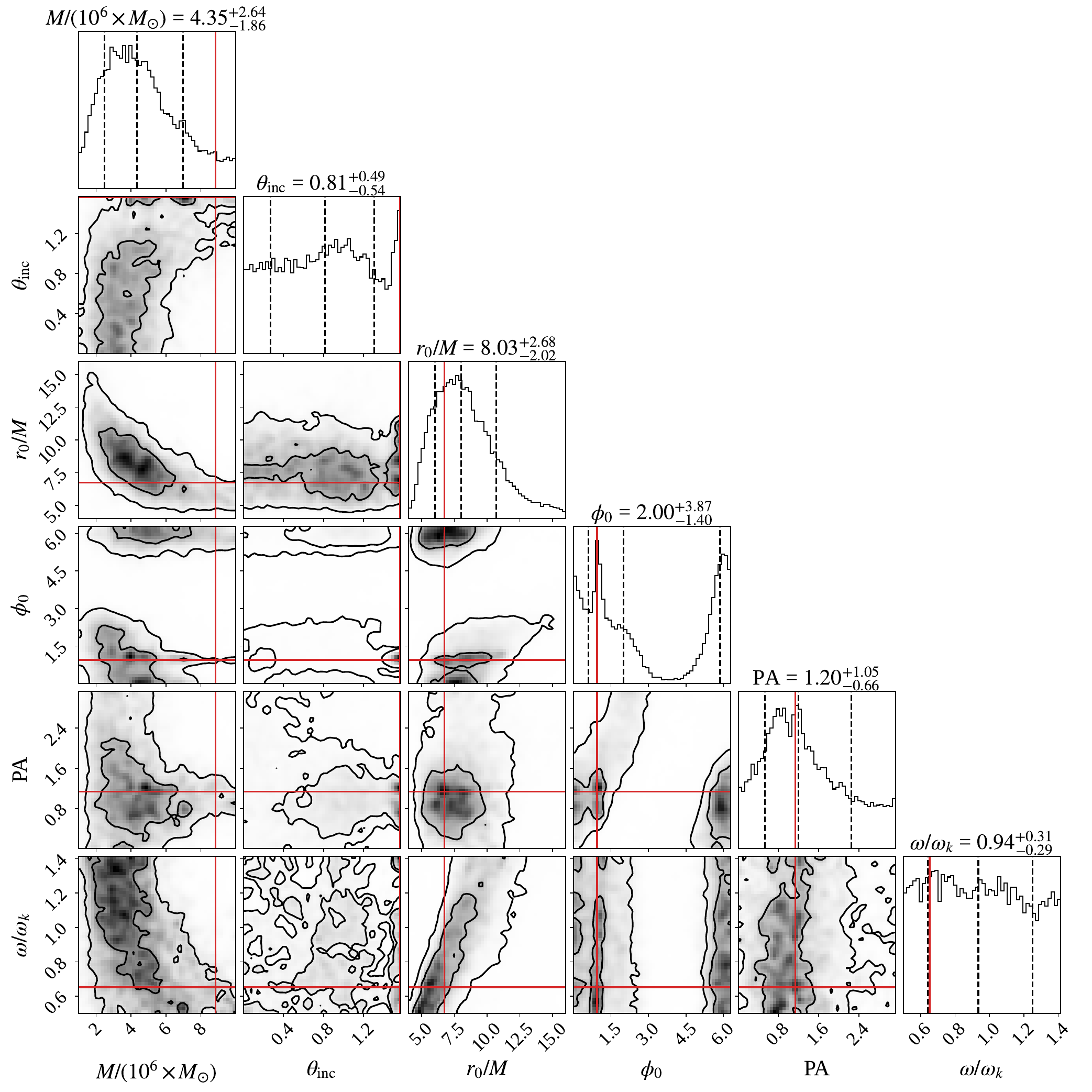} 
     \put(55,57){\includegraphics[width=0.44\textwidth]{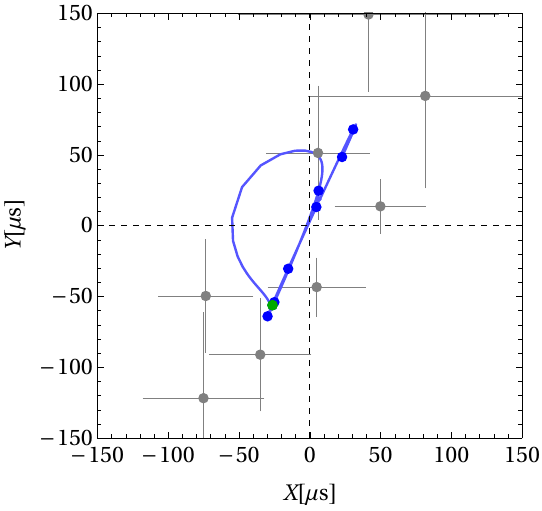}}
    \end{overpic}
    
  \end{subfigure}
  \hfill
  \begin{subfigure}[b]{0.49\textwidth}
  \caption{Jul 22, 2018}
    \begin{overpic}[width=\textwidth]{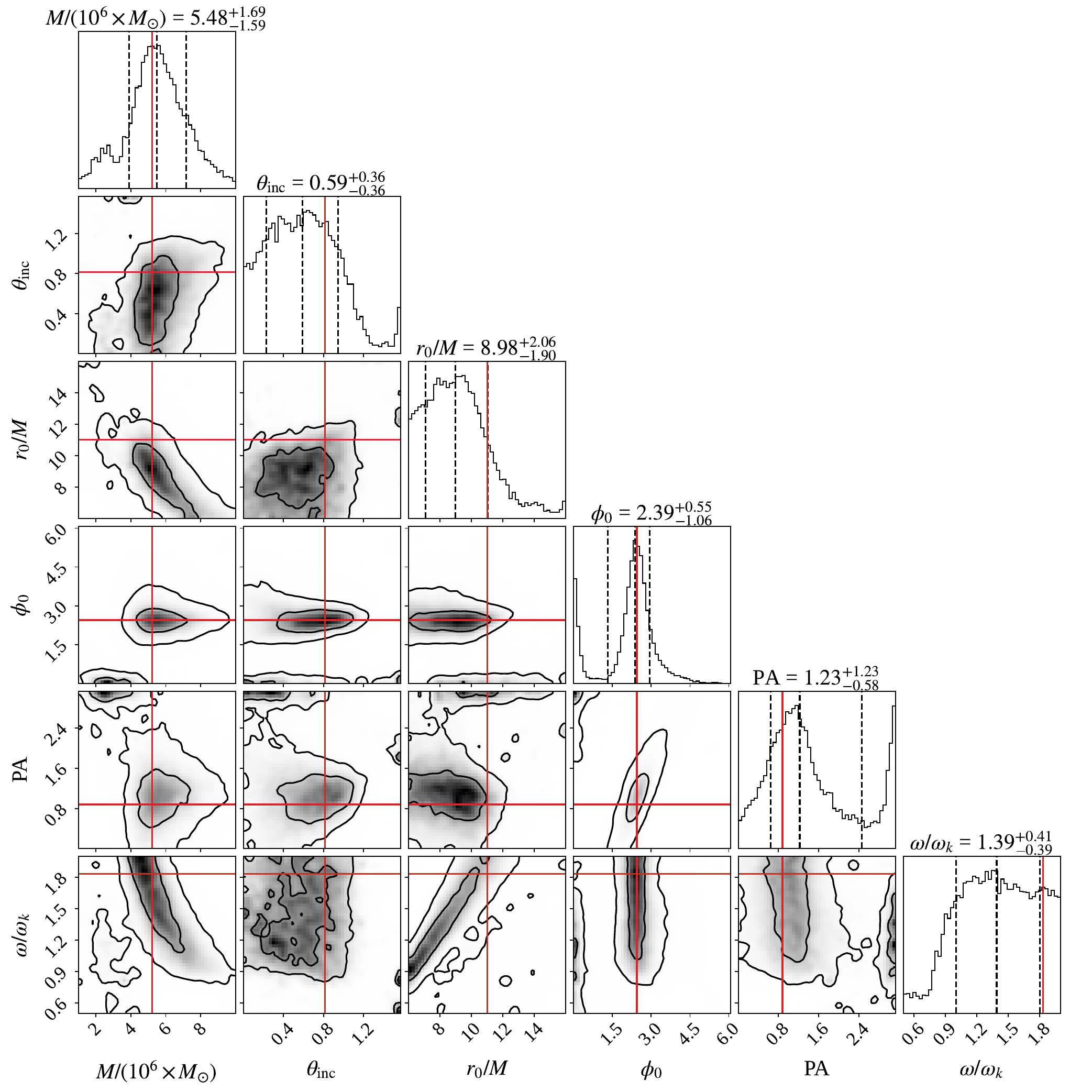} 
     \put(55,57){\includegraphics[width=0.44\textwidth]{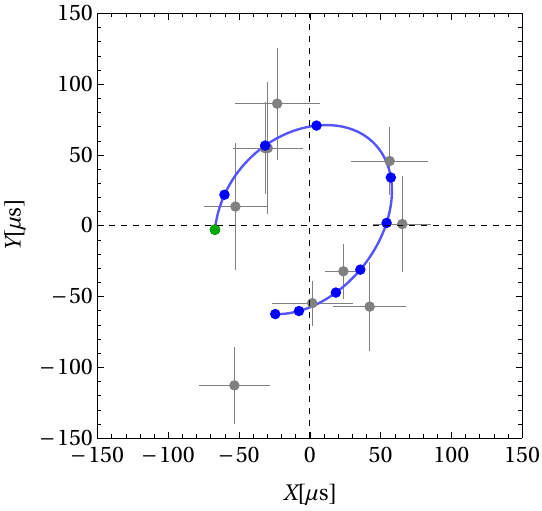}}
    \end{overpic}
  \end{subfigure}
  \vspace{1em}

  \begin{subfigure}[b]{0.49\textwidth}
  \caption{Jul 28, 2018}
    \begin{overpic}[width=\textwidth]{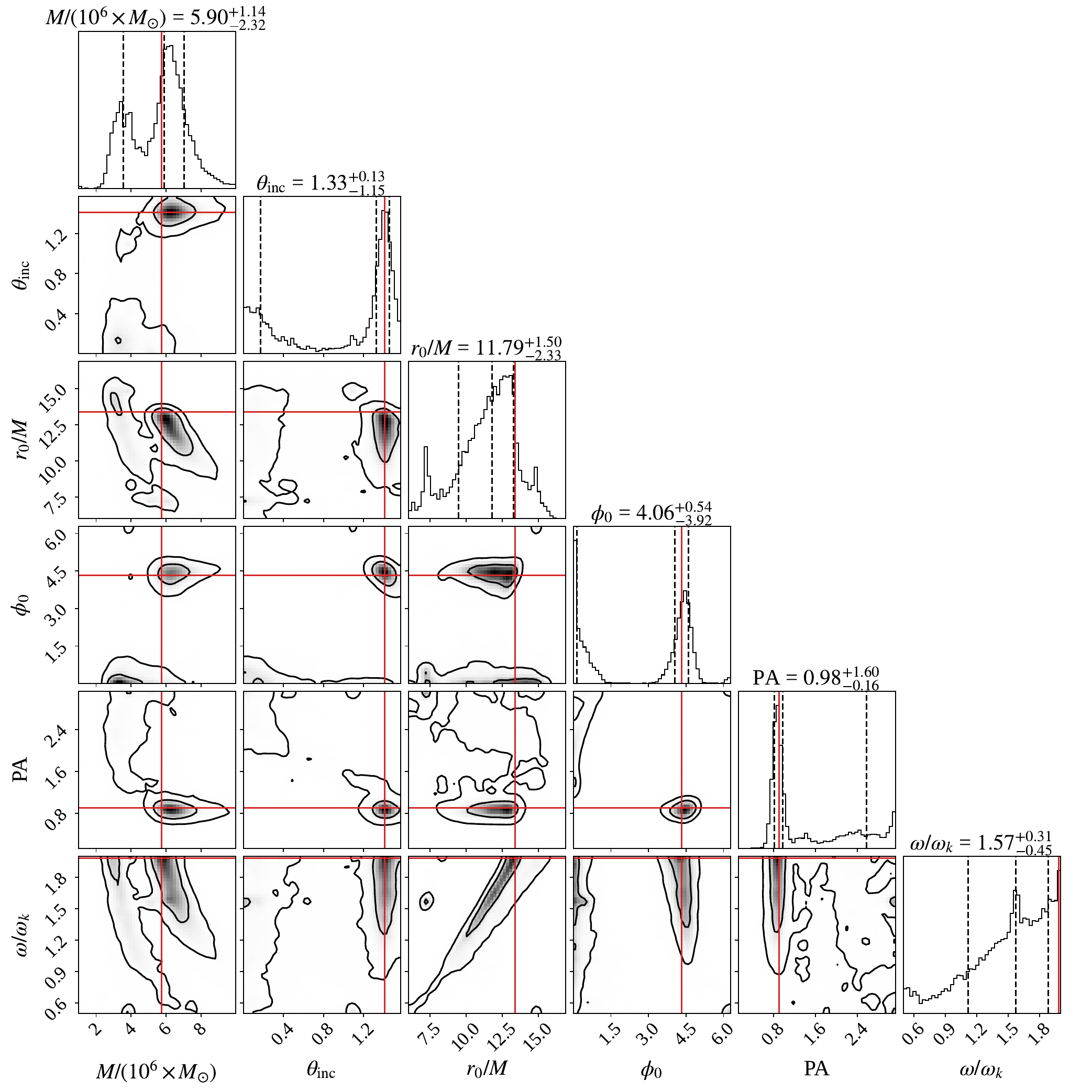} 
     \put(55,57){\includegraphics[width=0.44\textwidth]{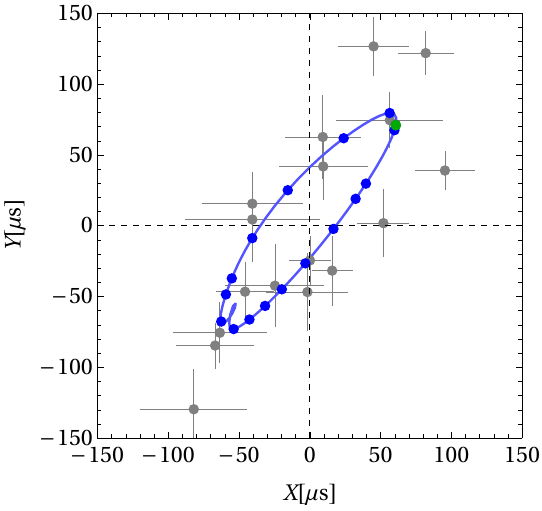}}
    \end{overpic}
  \end{subfigure}
  \hfill
  \begin{subfigure}[b]{0.49\textwidth}
  \caption{May 19, 2022}
    \begin{overpic}[width=\textwidth]{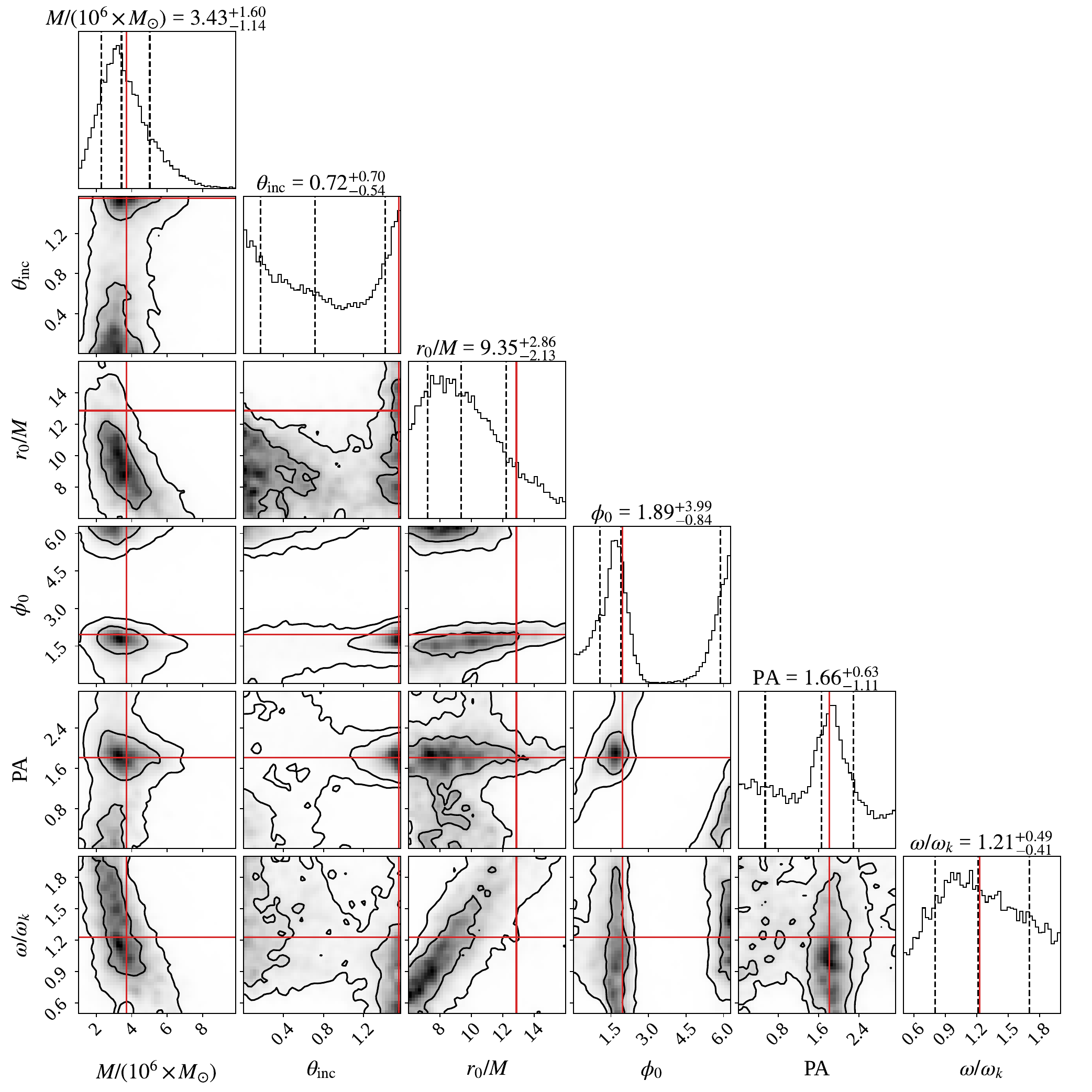} 
     \put(55,57){\includegraphics[width=0.44\textwidth]{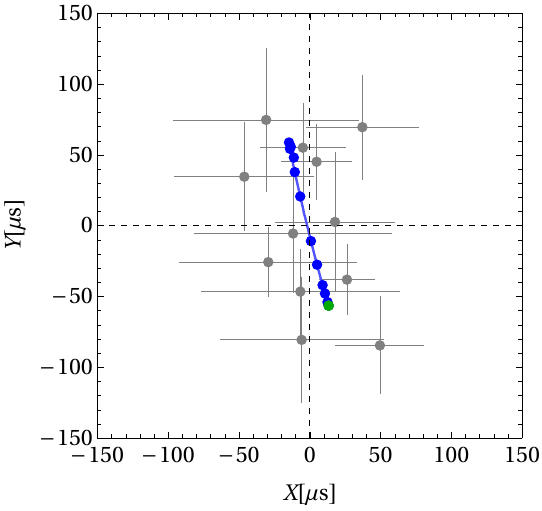}}
    \end{overpic}
    
  \end{subfigure}
  
    \caption{Bottom-left panels of (a)-(d):   posteriors for the four individual flare events. The red cross marks the globally best-fit parameters. 
    Top-right panels of (a)-(d): corresponding astrometry (gray points with error bars) and best-fit tracks (blue points) of a circularly orbiting hotspot for four individual flare events. Each panel corresponds to a flare event reported by the GRAVITY Collaboration. 
    From panels (a) to (d), the corresponding values of $\chi^2_\text{eff}$ are 1.32, 0.90, 3.61 and 0.69, respectively.
    }
    \label{diagram}
\end{figure}

While averaging data benefits from canceling systematic biases in the measurements, it might also smooth out the dynamical details of hotspot motion. In this part, we further investigated individual four flare events to examine the potential presence of the non-Keplerian motion. The posteriors for the four flare events are presented in bottom-left Fig.~\ref{diagram}. 
And in Tab.~\ref{T2}, we present the posterior median values of the parameters and corresponding uncertainties for the four flare events.

The four flare events yield results that deviate to some extent from those based on averaged data. The flare events on May 27, 2018 and May 19, 2022 indicate the inclination angle tending to be $\pi/2$, while the results of rest of flare events do not. It might originate from the systematic biases in the measurements, because the hotspots should be distributed within one single accretion disk.
From the posteriors for the individual flare events, there seems no significant evidence in favor of the non-Keplerian motions.  
And the global best-fit parameters deviate from the posterior median values.
It implies that the current data quality and quantity might be insufficient to draw a definitive conclusion regarding the presence of non-Keplerian motion. 

We plotted the apparent tracks of hotspots with the best-fit parameters for the four individual flare events, as shown in top-right panels of (a)-(d) in Figs.~\ref{diagram}. In chronological order, the $\chi^2_\text{eff}$ for the four flare events are 1.32, 0.90, 3.61 and 0.69, respectively.  
While the circular ortbit provides a reasonable fit for most flare events, we note that the goodness of fit varies across the four flares. In particular, the 19 May 2022 event yields a lower $\chi^2_\text{eff}$ despite having fewer and less precise data points than the 28 July 2018 event. This suggests that individual flare characteristics and data quality could influence the fitting performance, and some flare, like the event on 19 May 2022, might require more careful modeling.

\begin{table}
    \centering
        \caption{Posterior median values and uncertainties of the parameters modeled by hotspot circular orbits with the four individual astrometric data.}
    \begin{tabular}{c|cccc}
    \hline\hline
    Parameter & 27 May 2018 & 22 Jul 2018 & 28 Jul 2018 & 19 May 2022 \\
    \hline
    $M/(M_\odot\times10^6)$ & $4.35^{+2.64}_{-1.86}$   & $5.48^{+1.69}_{-1.59}$ & $5.90^{+1.14}_{-2.32}$ & $3.43^{+1.60}_{-1.14}$   \\
    $\theta_\text{inc}$          & $0.81^{+0.49}_{-0.54}$   & $0.59^{+0.36}_{-0.36}$ & $1.33^{+0.13}_{-1.15}$ & $0.72^{+0.70}_{-0.54}$   \\
    $r_0/M$                 & $8.03^{+2.68}_{-2.02}$   & $8.98^{+2.06}_{-1.90}$ & $11.79^{+1.50}_{-2.33}$& $9.35^{+2.86}_{-2.13}$   \\
    $\phi_0$                & $2.00^{+3.87}_{-1.40}$   & $2.39^{+0.55}_{-1.06}$ & $4.06^{+0.54}_{-3.92}$ & $1.89^{+3.99}_{-0.84}$   \\
    PA                      & $1.20^{+1.05}_{-0.66}$   & $1.23^{+1.23}_{-0.58}$ & $0.98^{+1.60}_{-0.16}$ & $1.66^{+0.63}_{-1.11}$   \\
    $\omega/\omega_k$       & $0.94^{+0.31}_{-0.29}$   & $1.39^{+0.41}_{-0.39}$ & $1.57^{+0.31}_{-0.45}$ & $1.21^{+0.49}_{-0.41}$   \\
    \hline
    \end{tabular}
    \label{T2}
\end{table}


\subsection{Hotspots in planar geodesic orbits}

\begin{figure}
    \centering
    \begin{overpic}[width=0.85\textwidth]{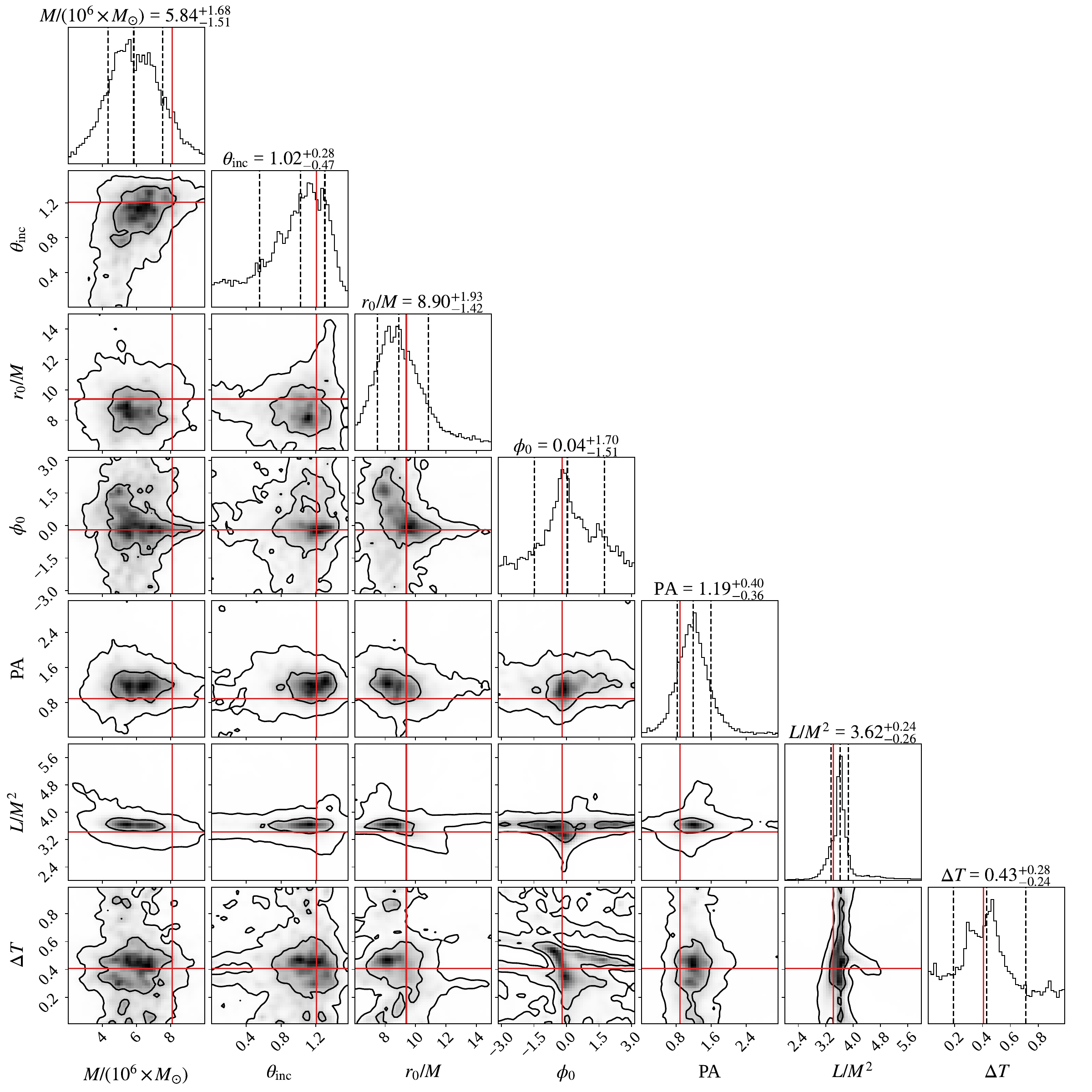} 
     \put(55,52){\includegraphics[width=0.40\textwidth]{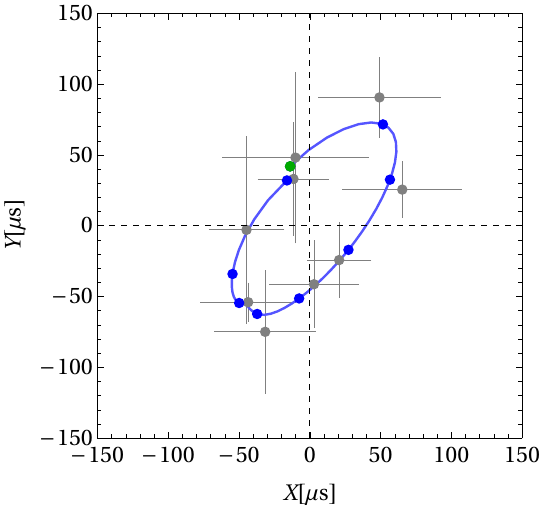}}
    \end{overpic} 
    \caption{Bottom-left panel:  the posteriors of model parameters obtained from MCMC sampling for the planar geodesic orbit model.  
    The red cross marks the globally best-fit parameters.
Top-right panel: observed flare
centroids (gray points) and the best-fit track (blue points) of the orbiting hotspot in the equatorial plane.  The value of $\chi^2_\text{eff}$ is 0.14. \label{non-circular orbits}}
\end{figure}

In order to examine the possibility that the non-Keplerian indication could deviate from the non-circular motion, we further fit the flare motions using non-circular geodesic orbits confined to the accretion disks. 
Fig.~\ref{non-circular orbits} presents the posteriors for this model. There is no strong negative correlation between the black hole mass and the orbital radius.
With the $\chi_\text{eff}^2=0.14$, the fit remains as good as  that of the circular orbit case. In Tab.~\ref{T3}, we present the posterior median values and uncertainties for the planar geodesic orbits.

\begin{figure}
    \centering
    \includegraphics[width=0.55\linewidth]{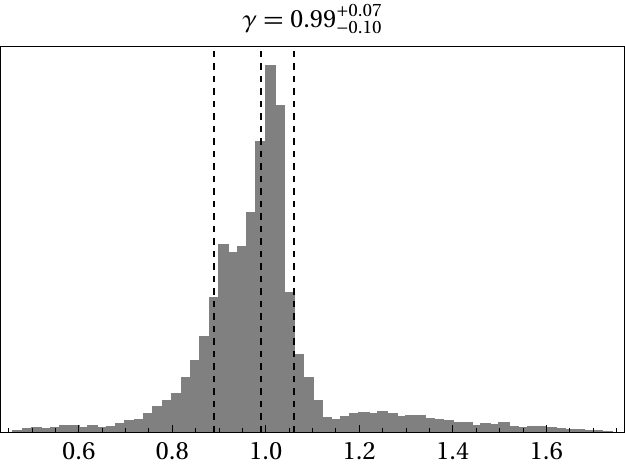}
    \caption{Posterior distribution of the circularity parameter $\gamma$.}
    \label{post}
\end{figure}

Because planar geodesic orbits are a generalization of circular Keplerian orbits, we can introduce a circularity parameter to quantify the derviation of non-circular orbits, namely, 
\begin{equation}
    \gamma \equiv L\sqrt{\frac{2 f(r_0)- r_0 f'(r_0)}{r^3_0 f'(r_0)}}~.
\end{equation}
Because we have $L=\sqrt{r^3_0 f'(r_0)/(2 f(r_0)- r_0 f'(r_0))}$ for circular orbits,  $\gamma=1$ represents no deviation. Using the posteriors of the parameters in Fig.~\ref{non-circular orbits}, we obtain the posterior distribution of the parameter $\gamma$ shown in Fig.~\ref{post}. Our results suggest that circular orbits are preferred among planar geodesic orbits.
\begin{table}
    \centering
        \caption{Posterior median values and uncertainties of the parameters modeled by hotspot in planar geodesic orbits with the avaraged astrometric data.}
    \begin{tabular}{c|ccccccc}
         \hline\hline
         Parameter & $M/(M_\odot\times10^6)$ & $\theta_\text{inc}$ & $r_0/M$ & $\phi_0$ & PA & $L$ & $\Delta T$ \\
         \hline
         --- & $5.84^{+1.68}_{-1.51}$ & $1.02^{+0.28}_{-0.47}$ & $8.90^{+1.93}_{-1.42}$ & $0.04^{+1.70}_{-1.51}$         & $1.19^{+0.40}_{-0.36}$ & $3.62^{+0.24}_{-0.26}$ & $0.43^{+0.28}_{-0.24}$  \\
         \hline
    \end{tabular}
    \label{T3}
\end{table}



\subsection{Error ellipses in astrometric data}
Because of the non-uniform u-v baseline coverage in interferometer measurements \cite{GRAVITYFirstlight}, the astrometric data $X_i$ and $Y_j$ could be correlated. 
This correlation produces error ellipses in the astrometric data, which can lead to the likelihood function in the form of
\begin{eqnarray}
    \mathcal{L}(\Theta_\text{model})=-\frac{1}{2}\sum_{i=1}^N\Bigg(\left(\frac{X_i-X_{\text{model}}(t_i)}{\sigma_{X_i}}\right)^2+\left(\frac{Y_i-Y_{\text{model}}(t_i)}{\sigma _{Y_i}}\right)^2\nonumber \\ +\frac{2\rho_i (X_i-X_{\text{model}}(t_i))(Y_i-Y_{\text{model}}(t_i))}{\sigma_{X_i}\sigma_{Y_i}}\Bigg) ~,   \label{like2}
\end{eqnarray}
where $\rho_i$ is correlation coefficient. The likelihood function in Eq.~(\ref{like2}) assumes independence between time sequence events $i$ and $j$. This is justified because the phase measurement on each baseline determines the sky position at a given time, and the non-uniform $(u,v)$-coverage \cite{GRAVITYFirstlight} does not introduce the temporal correlations.

\begin{figure}
    \centering
      \begin{subfigure}[b]{0.49\textwidth}
      \caption{$\rho_i=0.9$}
        \begin{overpic}[width=\textwidth]{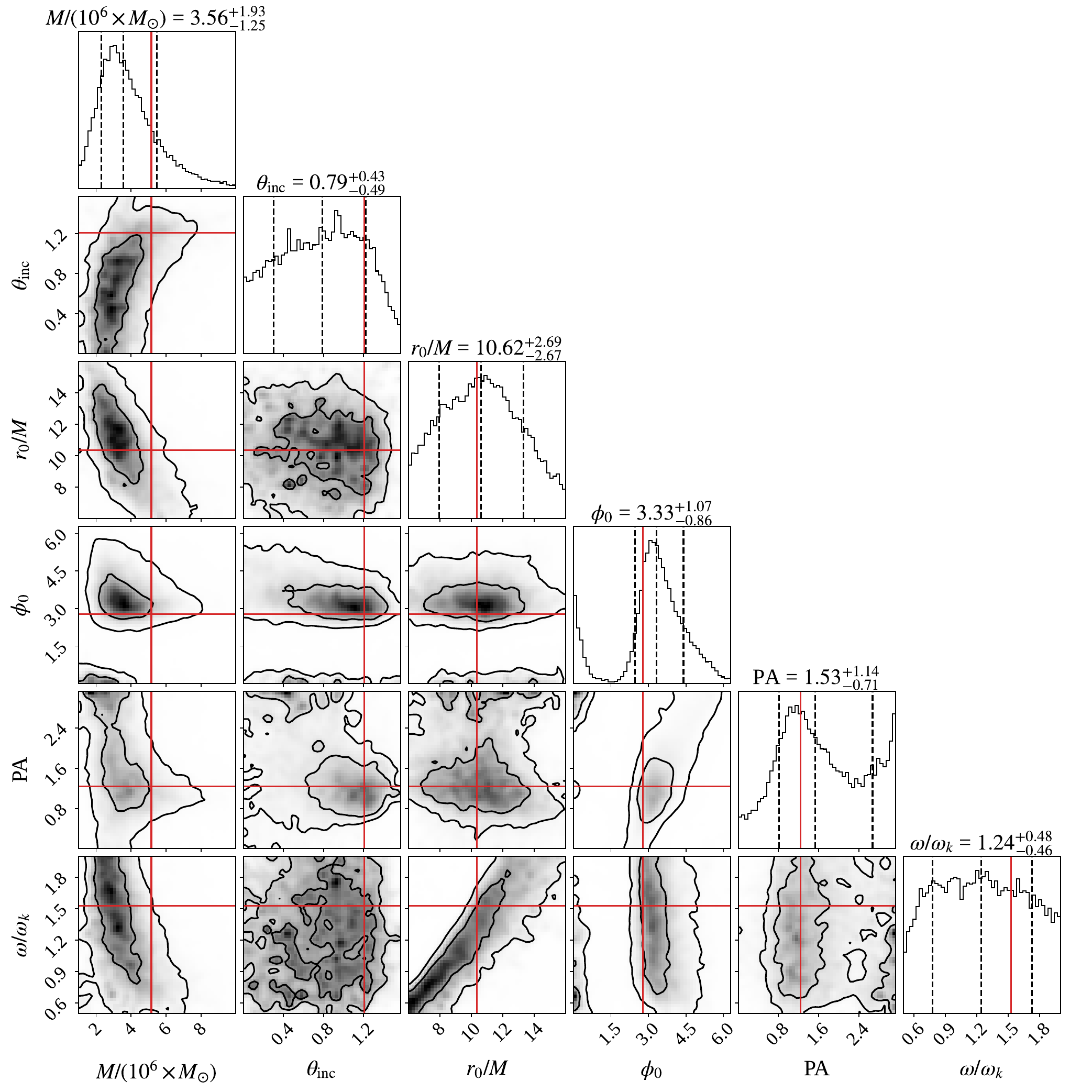} 
     \put(55,57){\includegraphics[width=0.44\textwidth]{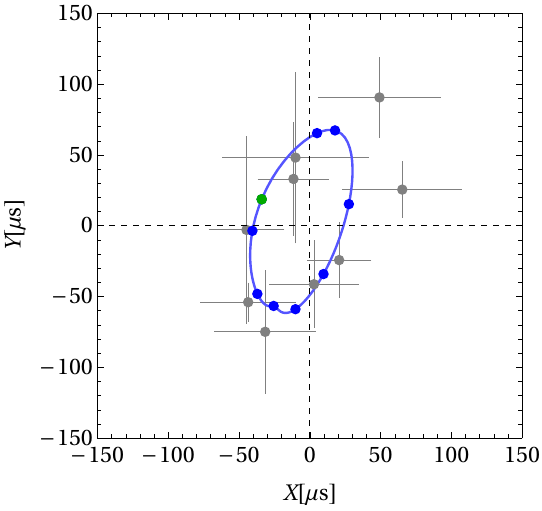}}
    \end{overpic}
    
  \end{subfigure}
  \hfill
  \begin{subfigure}[b]{0.49\textwidth}
  \caption{$\rho_i=-0.9$}
    \begin{overpic}[width=\textwidth]{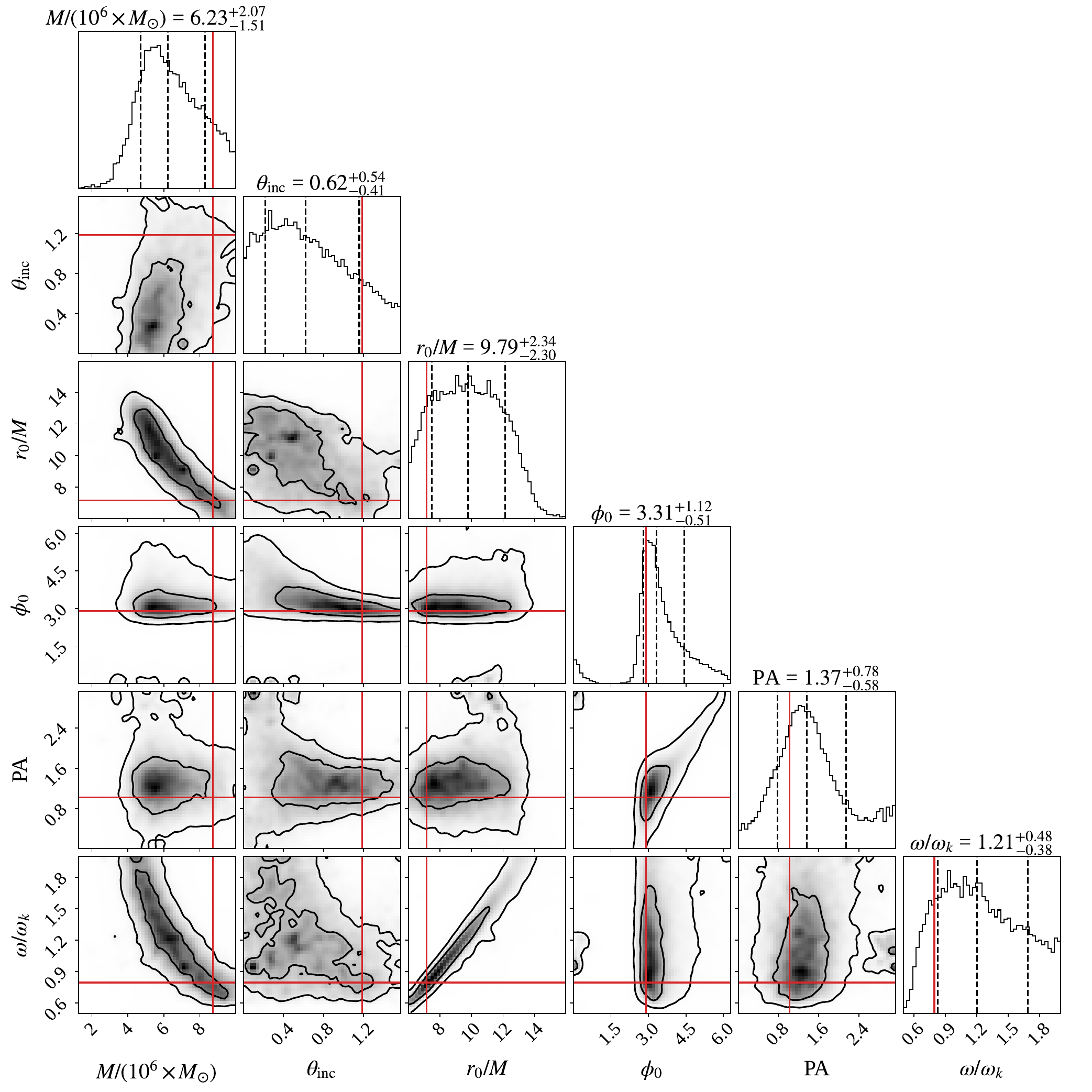} 
     \put(55,57){\includegraphics[width=0.44\textwidth]{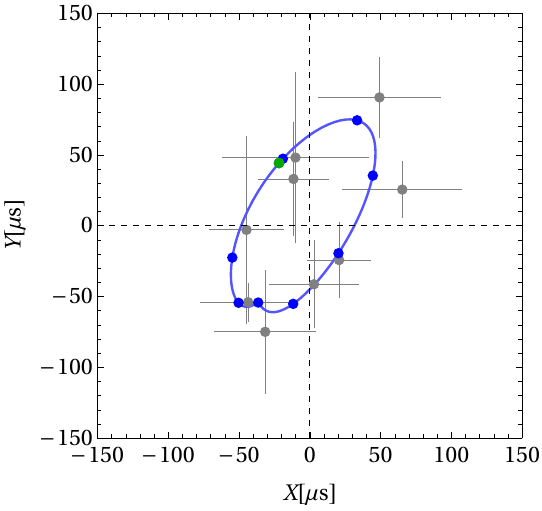}}
    \end{overpic}
  \end{subfigure}
  \vspace{1em}
  
    \caption{
        Bottom-left panel of (a) and (b): posteriors of model parameters (circular orbit) by assuming positive correlation (a) and negative correlation (b) in the averaged astrometric data. The red cross marks the globally best-fit parameters. Top-right panels of (a) and (b): corresponding astrometry (gray points with error bars) and best-fit tracks (blue points) of a circularly orbiting hotspot.    }
    \label{diagramRho}
\end{figure}

\begin{figure}
    \centering 
    \begin{overpic}[width=0.85\textwidth]{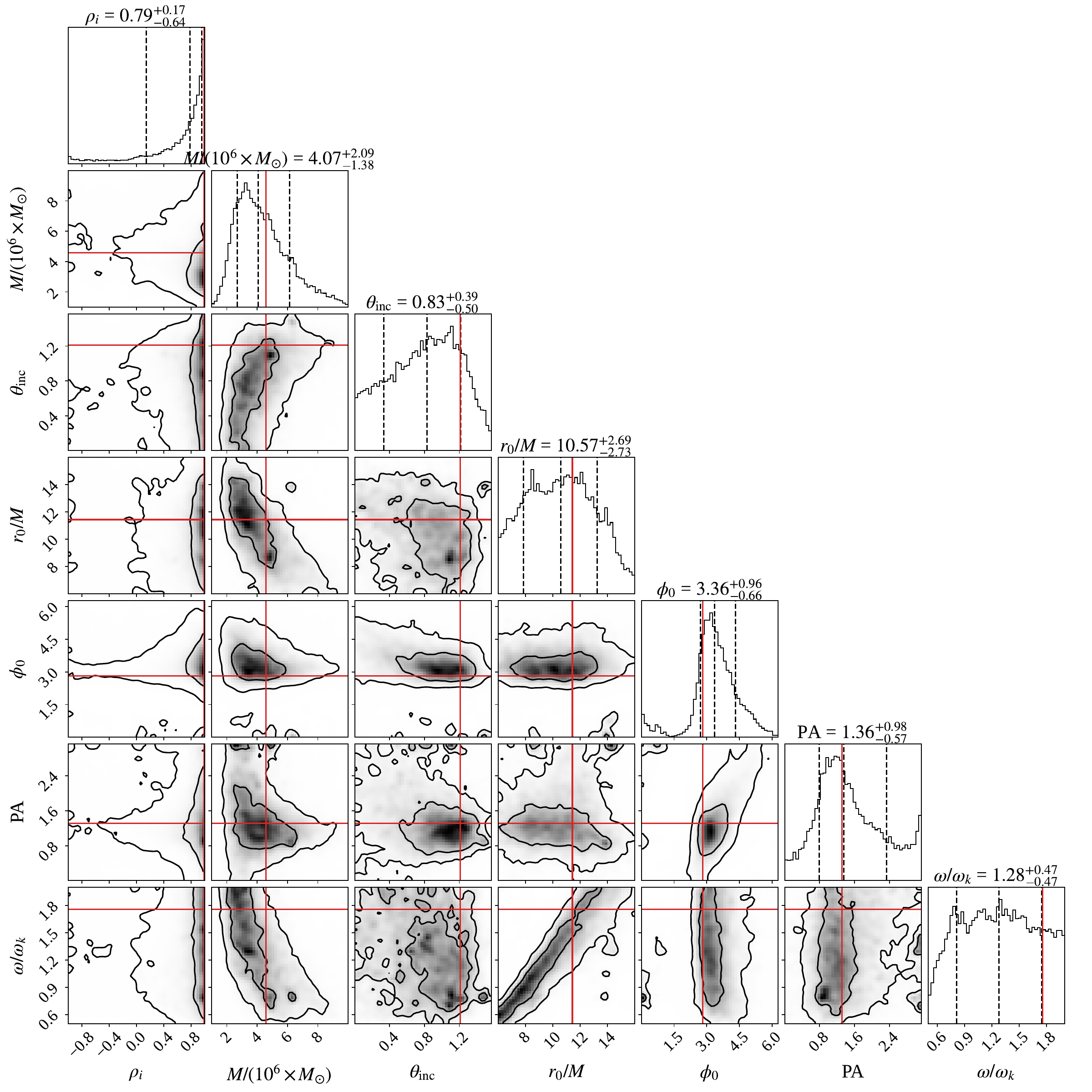} 
     \put(60,57){\includegraphics[width=0.37\textwidth]{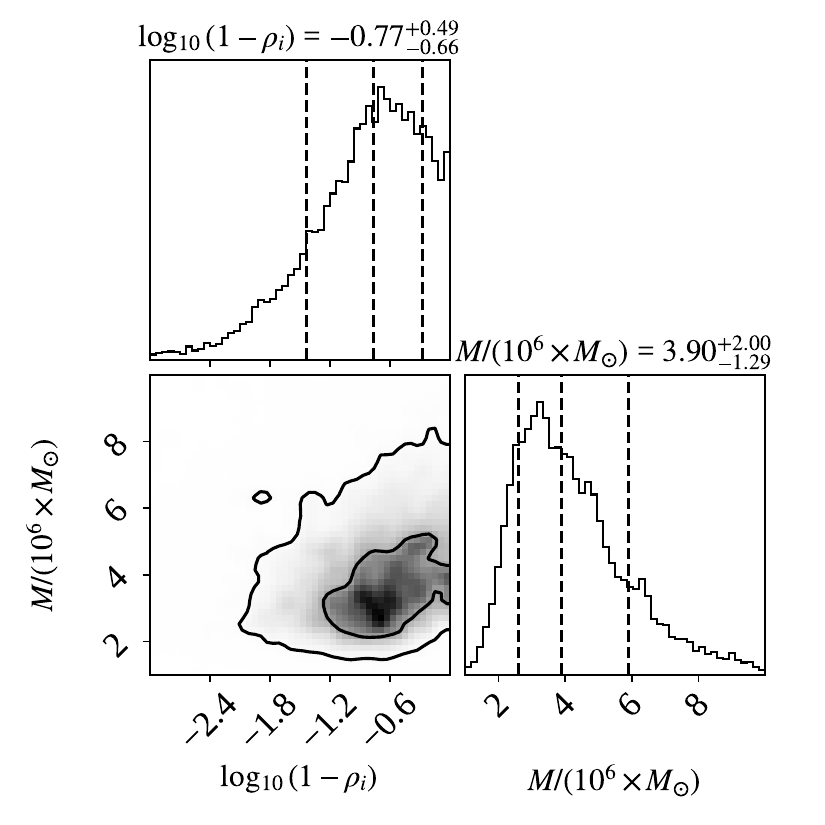}}
    \end{overpic} 
    \caption{Bottom-left panel:  posteriors of model parameters (circular orbit) with correlation coefficient $\rho_i$ treated as a free parameter.  
    The red cross marks the globally best-fit parameters.
Top-right panel: the rearranged posteriors for correlation coefficient and black hole mass. To show the posteriors in the regime of $\rho_i\rightarrow1$, we exclude the MCMC samples with negative $\rho_i$.  \label{F9}}
\end{figure}

In this part, we will study the influence from the correlated data on the parameter estimation. We consider strong correlation by setting $\rho_i=0.9$ and $-0.9$ to perform the parameter estimation, separately. In Fig.~\ref{diagramRho}, we present posteriors by fitting circular orbital hotspot with the averaged astrometric data. The results show that correlations in the astrometric data affect the posteriors, particularly that of the black hole mass. As shown in the left panel of Fig.~\ref{diagramRho}, the negative correlation ($\rho_i=-0.9$) produces a larger median mass, $M=6.23^{+2.07}_{-1.51} \times10^6 M_\odot$, than the established value \cite{2017:GReffectS2,GRAVITY:2021xju}, whereas the positive correlation ($\rho_i=0.9$) yields a smaller value of $M=3.56^{+1.93}_{-1.25} \times10^6 M_\odot$. In these fits, the non-Keplerian parameters are not well constrained. The preferred value of the correlation coefficient  can be determined by performing a parameter estimation in which $\rho_i$ is treated as a free parameter. We present the result in Fig.~\ref{F9}, which indicates a strong positive correlation in the astrometric data. This might originate from the interferometer telescope network being arranged in an "L"-like configuration \cite{GRAVITYFirstlight}. Here, the $\chi_\text{eff}$ is no longer a valid measure of the goodness of fit, because Eq.~(\ref{chieff}) does not hold for the likelihood function in Eq.~(\ref{like2}). We summarize the posterior median values and uncertainties for parameters in Tab.~\ref{TA}. 
\begin{table}
    \centering
        \caption{Posterior median values and uncertainties modeled by hotspot circular orbits with the correlation in averaged astrometric data.}
    \begin{tabular}{c|ccccccc}
    \hline\hline
         $\rho_i$  & $M/(10^6 \times M_\odot)$ & $\theta_\text{inc}$   & $r_0/M$   & $\phi_0$   & PA & $\omega/\omega_k$ \\
         \hline
         $0.9$  & $3.56^{+1.93}_{-1.25}$    & $0.79^{+0.43}_{-0.49}$   & $10.62^{+2.69}_{-2.67}$                              & $3.33^{+1.07}_{-0.86}$    & $1.53^{+1.14}_{-0.71}$   & $1.24^{+0.48}_{-0.46}$  \\
         $-0.9$ & $6.23^{+2.07}_{-1.51}$                       & $0.62^{+0.54}_{-0.41}$   & $9.79^{+2.34}_{-2.30}$                             & $3.31^{+1.12}_{-0.51}$    & $1.37^{+0.78}_{-0.58}$   & $1.21^{+0.48}_{-0.38}$  \\
         $0.79_{-0.64}^{+0.17}$ & $4.07_{-1.38}^{+2.09}$ & $0.83_{-0.50}^{+0.39}$ & $10.57_{-2.73}^{+2.69}$ & $3.36_{-0.66}^{+0.96}$ & $1.36_{-0.57}^{+0.98}$ & $1.28_{-0.47}^{+0.47}$
         \\
         \hline
    \end{tabular}
    \label{TA}
\end{table}

\


\section{Conclusions and discussions}\label{conclusion}

In this study, we used the flare data published by the GRAVITY Collaboration in 2023 \cite{abuter2023polarimetry} to investigate the possibility that the flare events exhibit deviations from circular Keplerian motion. For the averaged flare data, when the black hole mass was treated as a free parameter, the resulting hotspot tracks were consistent with circular Keplerian orbits, where $\omega/\omega_k=1.13^{+0.54}_{-0.42}$, and aligned with the previous study \cite{abuter2023polarimetry}. When the black hole mass was fixed at the canonical value of $M=4.3 \times 10^6 M_\odot$, the fitted tracks tended to favor the super-Keplerian motion at near 1$\sigma$ confidence level, where $\omega/\omega_k=1.45^{+0.35}_{-0.38}$.

While averaging data benefits from canceling systematic biases in the measurements, it might also smooth out the dynamical details of hotspot motion.  Thus, we analyzed four individual flare events, separately. In addition, we also explored planar geodesic motion models in an attempt to improve the fits. 
We summarize the reduced $\chi^2_\text{eff}$ and deviation parameters for hotspots in these orbits in Tab.~\ref{T4}, based on error estimation under the Bayesian framework.
For the individual flare data, there seems to be no significant evidence in favor of the non-Keplerian motions. Here, we have ignored the results for the flare on 28 Jul 2018, because of the poorly constrained posterior distribution of $\omega/\omega_k$.  We also show that the planar geodesic orbits also did not yield a significantly better fit compared to the circular orbits. The orbital circularity parameter is constrained to $\gamma=0.99^{+0.07}_{-0.10}$. 

\begin{figure}
    \includegraphics[width=0.9\textwidth]{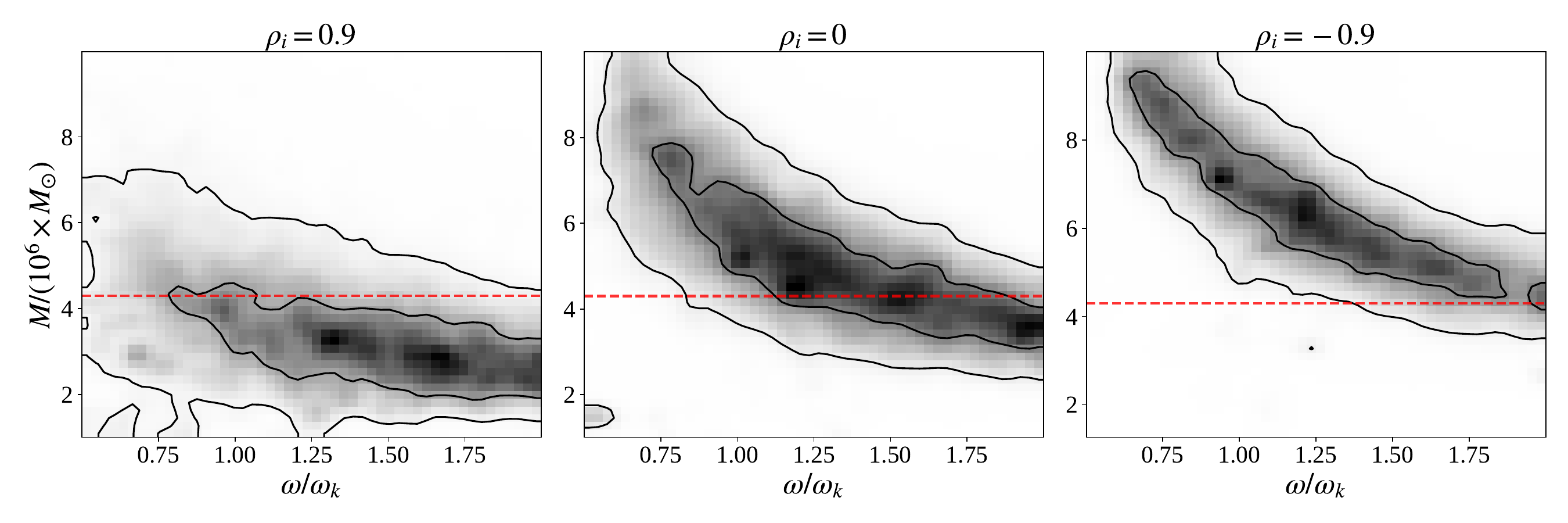}
    \caption{Joint posteriors for black hole mass and non-Keplerian parameter with given correlation coefficients $\rho_i$. The red dashed line represents $M=4.3\times 10^6 M_\odot$. \label{FL}}
\end{figure}

Our results, presented in Fig.~\ref{FL}, reveal a negative correlation between the black hole mass and the non-Keplerian parameter. Our models could not independently constrain both values. Namely, a higher inferred mass leads to a lower value of $\omega/\omega_k$, and vice versa.
Using the averaged data ($\rho_i=0$) and fixing the black hole mass to the established value of $M=4.3\times 10^6M_\odot$, the $\omega/\omega_k$ peaks at approximately $1.5$ as shown in the middle panel of Fig.~\ref{FL}. This initially suggests indication of the non-Keplerian motion, which is also supported by the posteriors shown in Figs.~\ref{23avg_fixed_mass-diagram_} and \ref{23avg_fixed prior-diagram_}. However, the statistical significance of this result is insufficiently high (approximately 1$\sigma$), and the unexpected negative correlation also requires careful interpretation. A comparison of the subplots in Fig.~\ref{FL} shows that the strength of this negative correlation changes depending on whether we account for correlations in the astrometric data. Specifically, for a fixed black hole mass, the inferred value of $\omega/\omega_k$ is sensitive to the correlation coefficients. This sensitivity suggests that the apparent indication of non-Keplerian motion is likely not physical. Instead, it probably originates from unaccounted-for uncertainties or correlations within the data itself rather than a discovery of new physics.

Compared with previous studies \cite{Fittinglightcurves,yfantis2024hot}, our results showed lower reduced $\chi^2_{\rm eff}$ values while exhibiting approximately ten times larger uncertainties for the non-Keplerian parameter $\omega/\omega_k$, which might be attributed to their use of polarimetric flare data. When polarimetry is considered, the polarization loops and temporal fluxes impose a strong constraint on the inclination angle, favoring a face-on orbits of the hotspots \cite{abuter2023polarimetry}. Namely, it yields a strong prior on the allowed incliantion angle in their astrometric fits. 
Since we showed that the astrometric data alone disfavor a moderate inclination (Figs.~\ref{23avg_diagram_} and \ref{non-circular orbits}), the combined astrometric and polarimetric data produces a better overall fit at the expense of a worse fit to the astrometry.
This explains the difference in our astrometric fit. Due to the large uncertainties in the astrometric fit alone, it seems reasonable to determine the inclination angle with polarimetry. In this sense, astrometric flare data alone might be insufficient to definitively determine whether the flares generally exhibit non-Keplerian motion. 



\begin{table}
    \centering
    \caption{Deviation parameters from circular Keplerian motion and effective $\chi_\text{eff}$ from the best fits \label{T4}}
    \begin{tabular}{cc|ccc}
    \hline\hline
    Model&data                               &  $\omega/\omega_k$        & $\gamma$ &  $\chi^2_\text{eff}$\\
    \hline
    Circular orbit&Averaged data                           &  $1.13^{+0.54}_{-0.42}$   &---&  0.18\\
    Circular orbit (fixed mass)&Averaged data                         &  $1.45^{+0.35}_{-0.38}$   &---&  0.29 \\
    Circular orbit&27 May 2018                                         &  $0.94^{+0.31}_{-0.29}$   &---&  1.32\\
    Circular orbit&22 Jul 2018                                         &  $1.39^{+0.41}_{-0.39}$   &---&  0.90\\
    Circular orbit&28 Jul 2018                                         &  $1.57^{+0.31}_{-0.45}$   &---&  3.61\\
    Circular orbit&19 May 2022                                        &  $1.21^{+0.49}_{-0.41}$   &--- &  0.69\\
    Non-circular geodesic orbit &Averaged data                               &  ---                      & $0.99^{+0.07}_{-0.10}$&  0.14\\
    \hline 
    \end{tabular}
    \label{tab:my_label}
\end{table}



\smallskip 

{\it Acknowledgments.} This work has been supported by the National Natural
Science Fund of China (Grants No. 12275034, No. 12347101, and No. 12305073). 
\bibliography{ref}

\end{document}